\documentclass[12pt]{iopart}
\usepackage{graphicx}

\newcommand{\mgf}{{\cal B}}
\newcommand{\AVP}{\vec{\cal A}}

\newcommand{\mbfr}{{\mathbf{r}}}

\begin{document}

\title{Wave packet approach to transport in mesoscopic systems}

\author{Tobias Kramer$^{1,2}$, Christoph Kreisbeck$^1$, and Viktor Krueckl$^1$}

\address{$^1$ Institut f{\"u}r Theoretische Physik, Universit{\"a}t Regensburg, 93040 Regensburg, Germany}
\address{$^2$ Department of Physics, Harvard University, Cambridge, MA 02138, USA} 

\begin{abstract}
Wave packets provide a well established and versatile tool for studying time-dependent effects in molecular physics. Here, we demonstrate the application of wave packets to mesoscopic nanodevices at low temperatures. The electronic transport in the devices is expressed in terms of scattering and transmission coefficients, which are efficiently obtained by solving an initial value problem (IVP) using the time-dependent Schr\"odinger equation. The formulation as an IVP makes non-trivial device topologies accessible and by tuning the wave packet parameters one can extract the scattering properties for a large range of energies.
\end{abstract}

\section{Introduction.}

Low-dimensional mesoscopic devices are systems in which electronic movement is confined along a two-dimensional layer within the semiconductor \cite{Davies1998a}. Molecular-beam epitaxy allows one to produce extremely clean systems where the free path length of electrons at very low temperature reaches from the $\mu$m up to the mm scale before inelastic collisions occur and coherence is lost \cite{Ferry1997a}. The large coherence length requires to treat the system on a quantum-mechanical level since interference and diffraction effects occur. The electron densities in the electron layer are on the order of $10^{15}$~m$^{-2}$, with a positively charged background of approximately the same density located at a distance of $30-100$~nm from the electronic layer. The electrons are occupying the available states up to the Fermi energy $E_F$, which can be adjusted by external gates. Experimentally, the properties of the system are probed by alloyed metallic contacts which reach down to the electronic layer. At low temperatures the current through the system is measured as a function of the applied voltages at the contacts. In principle, the electric circuit is closed through a battery, where charges are separated and after travelling through the wires get injected in the semiconductor \cite{Poetz1989a}. A quantum-mechanical description of the complete circuit cannot be achieved. Instead one implements an open system, where the contacts inject the electrons coming from an external reservoir, which can be viewed as an environment. The determination of the correct boundary conditions at the contacts is a considerable challenge and presents a still unsolved problem for interacting particles.
Theoretically, transport in mesoscopic systems at low temperatures and at low currents is often described within non-interacting models. The justification for this drastic simplification lies in the idea that only electrons with energies close to the Fermi energy are participating in the net transport and that Pauli blocking suppresses scattering events between electrons below and above the Fermi energy completely. Considerable effort has to be made to correctly determine the effective potential landscape for the electrons at the Fermi energy, since the positive background charges are largely screened by the electrons within the device. The effect of screening has to be calculated for a specific geometry in the presence of metallic gates and randomly distributed donor charges \cite{Davies1998a,Aidala2007a}.  The resulting  mean-field approach to mesoscopic transport has worked reasonably well for a wide variety of systems, but interactions effects are known to play an important role in small devices and in magnetic fields, where charging effects (Coulomb blockade) and correlations (fractional quantum Hall effect) occur.

For predicting the current through the system in response to a small change of voltage at one of the contacts, quantitative calculations rely on knowing the transmission (or scattering) probability from the source contact to the drain contact. The irregular device geometry necessitates the use of numerical methods to evaluate the scattering cross-sections. Scattering processes can be described in a stationary (time-independent) way using eigenstates of the asymptotic part of the system or using a time-dependent approach based on the propagation of wave packets. Here, we will discuss the time-dependent wave packet method, which allows us in principle to handle both, non-interacting and interacting systems. The wave packet techniques used in atomic and molecular physics have to be changed in several ways in order to be applicable for nanodevices \cite{Kramer2008a}. We will first give a detailed discussion of the non-interacting theory and in the last section comment on interaction effects.

\section{Device layout and the asymptotic region}

The proper definition of a scattering matrix requires to introduce asymptotic regions where the scattering potential is absent and exact eigenstates can be constructed. In the two-dimensional systems under consideration, the contacts inject electrons into waveguide like structures which guide the electrons to the scattering region. If we consider the contacts as emitting \textit{incoherently} electrons, we obtain the emitted electron current by convolution of the product of the local density of states (LDOS) at the point of injection with the group velocity and the occupation probability given by the Fermi-Dirac distribution.

\subsection{Parabolic waveguide in a homogeneous magnetic field}

\begin{figure}[t]
\includegraphics[width=0.5\textwidth]{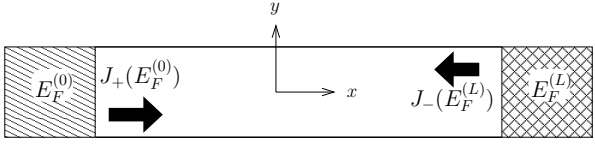}
\caption{Current flow in a wave guide coupled to two reservoirs with Fermi energies $E_F^{(0)}$ and $E_F^{(L)}$. \label{fig:reservoir}}
\end{figure}

\begin{figure}[t]
\includegraphics[width=0.5\textwidth]{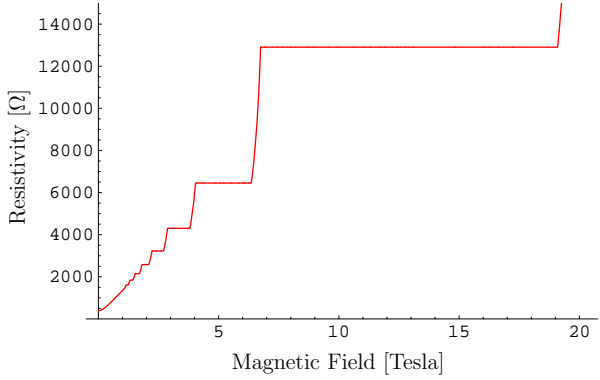}
\caption{Two-terminal resistivity $1/\sigma_{\rm net}$ as function of the magnetic field. The left reservoir is at $E_F^{(0)}=16.5$~meV, the right one at $E_F^{(L)}=17.5$~meV.\label{fig:twoterm}}
\end{figure}

An instructive example consists of an electron released into a parabolic potential $V(y)=\frac{1}{2}m\omega_y^2 y^2$ along the $y$-direction.
In addition a homogeneous magnetic field ${\cal B}$ is present along the $z$-axis together with a strongly confining potential along the $z$-directon, which quantizes the system along this axis and reduces the effective dimensionality of the subsystem under consideration to two dimensions. The Hamiltonian for the two-dimensional motion reads in the Landau gauge $\AVP=(-{\cal B} y,0,0)$ \cite{Johnson1983a}:
\begin{equation}\label{eq:HamiltonParab}
H=\frac{1}{2m}{(-\rmi\hbar\partial_x+q{\cal B}y)}^2-\frac{\hbar^2}{2m}\partial_y^2+\frac{1}{2}m \omega_y^2 y^2.
\end{equation}
Here, $m$ denotes an effective electronic mass (in AlGaAs/GaAs layers $m=0.067\;m_e$), which results from taking into account the periodic crystal structure of the semiconductor within the $\mathbf{k}\cdot \mathbf{p}$ expansion \cite{Davies1998a}.
It is useful to introduce the cyclotron frequency $\omega_C=\frac{e\mgf}{m}$ and to express $\omega_y$ in terms of another frequency 
$\Omega^2=\omega_y^2+\omega_C^2$.
The normalized eigenfunctions and eigenenergies of the Hamiltonian~(\ref{eq:HamiltonParab}) are:
\begin{eqnarray}
\phi_{n,k_x}(x,y)=\frac{\rme^{\rmi k_x x}}{\sqrt{2\pi}}\frac{1}{\sqrt{2^n n! \sqrt{\pi}l}}
\exp\left[-\frac{{(y-\frac{\omega_C}{\Omega^2}\frac{\hbar k_x}{m})}^2}{2 l^2}\right]
H_n\left(\frac{y-\frac{\omega_C}{\Omega^2}\frac{\hbar k_x}{m}}{l}\right),
\end{eqnarray}
\begin{equation}
l=\sqrt{\frac{\hbar}{m\Omega}},\quad E_{n,k_x}=\left(n+\frac{1}{2}\right)\hbar\Omega+\frac{\hbar^2 k_x^2}{2m}\frac{\Omega^2-\omega_C^2}{\Omega^2}.
\end{equation}
We express the LDOS in terms of a sum and an integral over the squared eigenfunctions weighted with the eigenenergies:
\begin{eqnarray}
n_{\rm parab}(\mbfr;E)
&=&\sum_{n=0}^{\infty}\int_{-\infty}^{\infty}\rmd k_x\,
\delta(E-E_{n,k_x}) {|\phi_{n,k_x}(x,y)|}^2\nonumber\\
&=&
\sum_{n=0}^{\infty}\sum_{k=k_-,k_+} 
{\left|\frac{\partial E_{n,k_x}}{\partial k_x}\right|}^{-1}_{k_x=k}
\Theta(E-\hbar\Omega(n+1/2)) {|\phi_{n,k}(x,y)|}^2,
\end{eqnarray}
with
\begin{equation}
\frac{\partial E_{n,k_x}}{\partial k_x}=\frac{\hbar^2 k_x}{m}\frac{\Omega^2-\omega_C^2}{\Omega^2},
\quad
k_{\pm}=\pm\frac{\sqrt{2m [E-\hbar\Omega(n+1/2)]}\Omega}{\hbar\sqrt{\Omega^2-\omega_C^2}}.
\end{equation}
For each quantum number $n$, the energy integrated LDOS becomes
\begin{equation}\label{eq:leveldeg}
N_n=\int_{-\infty}^\infty\rmd E\;n_{{\rm parab},n}(\mbfr;E)
=\frac{\Omega^2}{\omega_C^2} \frac{e {\cal B} }{2\pi\hbar}.
\end{equation}
No spatial dependence of $N_n$ remains after the energy integration. The density of electrons per quantized level (the degeneracy) differs from the quantization in a purely magnetic field (or orthogonal electric and magnetic fields) by the factor $\Omega^2/\omega_C^2$ \cite{Kramer2003a}.\\
We proceed to calculate the quantum mechanical current emitted with fixed energy from the point $\mbfr$ \cite{Kramer2003d}. The drift current $j(\mbfr;E)$ is given by the product of the group velocity $\partial_{k} E_{n,k}/\hbar$ times the local density of states \cite{Beenakker1991a}. We obtain two counterpropagating components of the current:
\begin{eqnarray}
j_{x,\pm}(\mbfr;E)
&=&\frac{e}{\hbar}\sum_{n=0}^\infty\int\rmd k_x\; \delta [E-E_{n,k_x}] {|\phi_{n,k}(\mbfr)|}^2 \Theta(\pm k_x) \frac{\partial E_{n,k_x}}{\partial k_x}  \\\nonumber
&=&\pm\frac{e}{\hbar}\sum_{n=0}^\infty \Theta(E-\hbar\Omega(n+1/2))\;{|\phi_{n,k_\pm}(\mbfr)|}^2.
\end{eqnarray}
The total current through a closed surface around $\mbfr$ is given by the sum
$j(\mbfr;E)=|j_{x,+}(\mbfr;E)|+|j_{x,-}(\mbfr;E)|$. At energies $\hbar\Omega(n+1/2)$ the product of the singular LDOS with the group velocity results in a finite current. %
The energy integral over the current weighted with the occupation probability gives
\begin{eqnarray}\label{eq:jtotr}
J(\mbfr;E_F)
&=&  \int_{-\infty}^{\infty} \rmd E\; f(E,E_F,T)\; \left(|j_+(\mbfr;E)|+|j_-(\mbfr;E)|\right)\\\nonumber
&=&e \sum_{n=0}^{\infty} \int_{-\infty}^{\infty} f(E,E_F,T) 
\Theta(E-\hbar\Omega(n+1/2)) \times \\
&&\qquad\left[ {|\phi_{n,k_+}(x,y)|}^2 + {|\phi_{n,k_-}(x,y)|}^2\right] \rmd E\;.
\end{eqnarray}
where $f(E,E_F,T)=1/(\exp[(E-E_F)/(k_B T)]+1)$ denotes the Fermi-Dirac distribution.
We obtain the global current by an integration over the local current given in eq.~(\ref{eq:jtotr}):
\begin{eqnarray}
J(E_F)
&=&
\int_{0}^{L}\rmd x
\int_{-\infty}^{\infty} \rmd y\;
J(\mbfr;E_F)\nonumber\\
&=&
\frac{2 L e}{2\pi\hbar}
\sum_{n=0}^{\infty}\int_{-\infty}^{\infty}\rmd E\; f(E,E_F,T) \Theta(E-\hbar\Omega(n+1/2)).
\label{eq:jef}
\end{eqnarray}
From eq.~(\ref{eq:jef}), we calculate the current through a parabolic waveguide which is coupled at $x=0$ to electrons in a reservoir with Fermi energy $E_F[0,y]=E_F^{(0)}$ and at $x=L$ to a second reservoir with Fermi energy $E_F[W,y]=E_F^{(L)}$ (see fig.~\ref{fig:reservoir}). For simplicity, we set the temperature to zero.
The net current $J_{\rm net}$ along $x=0\ldots L$ is the sum of the counterpropagating currents from both reservoirs (each reservoir has an extension of $\Delta L$ along $x$): $J_+(E_F^{(0)})=J(E_F^{(0)})/2$ and $J_-(E_F^{(L)})=J(E_F^{(L)})/2$. For $ \hbar\Omega (M+1/2)<E_F^{(L)}<E_F^{(0)}<\hbar\Omega (M+3/2)$, $M=0,1,2,\ldots$
\begin{equation}\label{eq:currentmode}
J_{\rm net}=\frac{J(E_F^{(L)})-J(E_F^{(0)})}{2\Delta L}=\frac{e (M+1)}{2\pi\hbar}\left(E_F^{(0)}-E_F^{(L)}\right),
\end{equation}
where $M+1$ is the number of occupied modes below the Fermi energy. The net current depends only on the difference of the Fermi energies along the waveguide. Eq.~(\ref{eq:currentmode}) can be obtained without explicitly referring to the LDOS and the local drift current (see \cite{Beenakker1991a}, eq.~(12.9)), but the apparent detour using the LDOS makes the connection with the source approach to quantum transport \cite{Kramer2003d} more transparent. Since the squared eigenfunctions $|\phi_{n,k_x}|^2$ are normalized to $1/(2\pi)$, the spatial integration just gives a factor $\Delta L/(2\pi)$, while the partial derivative of $E_{n,k_x}$ cancels the inverse partial derivative coming from the energy integration of the DOS:
\begin{eqnarray}\nonumber
J_{\rm net}&=&e\sum_{n}\int\rmd k_x\;
 \int_{E_F^{(0)}}^{E_F^{(L)}}\rmd E\; 
\delta(E-E_{n,k_x})\left[\int\rmd\mbfr\;{|\phi_{n,k_x}(\mbfr)|}^2\right]
\frac{\partial E_{n,k_x}}{\hbar\partial k_x}
\\\label{eq:jnet}
&=&\frac{e\Delta L}{2\pi\hbar}\sum_{n}\int_{E_F^{(0)}}^{E_F^{(L)}}\rmd E\; \Theta(E-\hbar\Omega(n+1/2)).
\end{eqnarray}
Eq.~(\ref{eq:jef}) is converted to a conductivity by dividing $J_{\rm net}$ by the difference in Fermi energies:
\begin{equation}\label{eq:buttiker}
\sigma_{\rm net}=e\frac{J_{\rm net}}{E_F^{(0)}-E_F^{(L)}}=(M+1)\frac{e^2}{2\pi\hbar}
\end{equation}
The resulting conductivity is quantized in units of $e^2/h$, as long as the voltage difference between the two reservoirs does not exceed the mode separation $\hbar\Omega$. The resulting resistivity $\rho=1/\sigma_{\rm net}$ is shown in fig.~\ref{fig:twoterm} for a fixed Fermi energy as function of the magnetic field.  The origin for the quantization in steps of $e^2/h$ is not the quantized number of available states per Landau level (\ref{eq:leveldeg}), but rather the effective reduction to a longitudinal one-dimensional channel.

\subsection{Scattering and multiple channels}

\begin{figure}[t]
\includegraphics[width=0.65\textwidth]{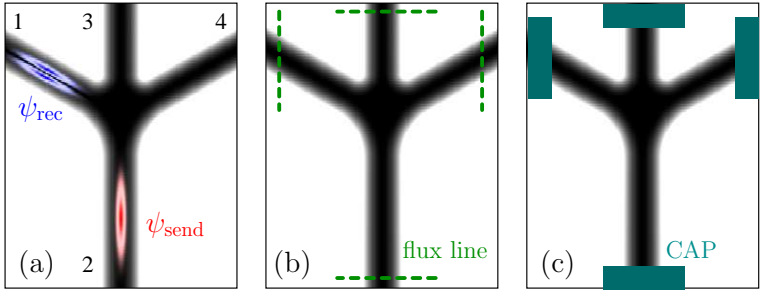}
\caption{Geometry of a four terminal device with arms 1 and 4 and main channels 2 and 3. The simulation area covers a width of $0.7\;\mu$m and a height of $0.9\;\mu$m.
(a) The sender wave packet sits in the lower arm and the receiver wave packet in the left arm in order to track the time-dependent correlation function. (b) Flux lines in the asymptotic regions are indicated, through which the time-dependent probability flux is recorded. (c) Complex absorbing potentials (CAPs) at the end of the channels mimic a semi-infinite extension and are used to record the flux leaving the simulation region.\label{fig:hallcross}}
\end{figure}

The conductance quantization in eq.~(\ref{eq:buttiker}) holds only in the absence of scattering within the waveguide. In the presence of scattering due to deformations and impurities in the waveguide, the conductivity will get reduced. Numerical methods are required to calculate the transmission and conductance of such a system. The attachment of additional asymptotic channels (terminals) to the waveguide shown in fig.~\ref{fig:hallcross} requires to extend the formalism to handle the multi-terminal case. This generalization was done by B{\"u}ttiker \cite{Buettiker1988a} and leads to the following expression for the current from channel $i$
\begin{equation}\label{eq:current}
I_i=\frac{e}{h}\int_{-\infty}^{\infty}\,\rmd E\ \sum_{j \neq i, n_i, n_j}
|t_{in_i\,jn_j}(E)|^2\left( f(E,\mu_i,T)-f(E,\mu_j,T)\right),
\end{equation}
where  $t_{in_i\,jn_j}$ denotes the transmission amplitude for scattering from the transverse mode $n_j$ in arm $j$ into the mode $n_i$ in arm $i$.

\section{From wave packets to transmissions}

The Landauer-B{\"u}ttiker formalism relates the elements of the scattering matrices with the currents and voltages in the system. The system
consists of a central region of irregular shape, which is connected to semi-infinite channels, which are assumed to be free of imperfections which cause scattering.
The semi-infinite channels allow one to introduce a set of channel eigenstates and to study transmission from a well-defined asymptotic state. A similar concept is behind Wigner's and Eisenbud's $\mathbf{R}$-matrix approach, where an artificial boundary is introduced at which the asymptotic regions and the interaction/scattering region get connected. Within the system the probability current is conserved and all equations can be derived by considering the probability flux through a closed surface. The probability flux describes stationary \cite{Moshinsky1951a} as well as time-dependent processes \cite{Moshinsky1951b}.
The time-dependent approach transforms the scattering problem into an initial value problem (IVP), which circumvents the explicit construction of eigenstates. The adventage of the IVP formulation become most apparent if transmission coefficients for a wide energy range are required (to study how the conductivity changes with Fermi energy, non-zero temperatures, or small voltages). In practice we construct a wave packet of finite extent by forming a superposition of plane waves along the waveguide with a specific transversal mode of the waveguide.

\section{Propagation methods}

An analytic time-evolution of the wave packet in the scattering region is only known for very few problems \cite{Grosche1998a}.
Thus in general numerical methods are required to solve the time-dependent Schr\"odinger equation for this part of the system.
Roughly the methods can be divided in two classes, (i) methods based on the Trotter expansion of the time-evolution operator and (ii) polynomial expansions of the time-evolution operator. The time evolution typically depends on a kinetic part $T=-\hbar^2\nabla^2/(2 m)$ and a potential part $V(\mbfr)$ which do not commute.
In the Trotter approach \cite{Feit1982a} the completete time interval is broken into $N$ small time-steps $\Delta t$ and
one expands the time-evolution operator into products of exponential functions for the kinetic and the potential part
\begin{equation}
U(t,t+N\Delta t) = \rme^{-\rmi (T+V) N \Delta t/\hbar} \approx
 \left ( \rme^{-\rmi V \Delta t/2\hbar}
 \rme^{-\rmi T \Delta t/\hbar}
 \rme^{-\rmi V \Delta t/2\hbar}
 \right )^N
.
\end{equation}
The commutation error of potential and kinetic energy in this expansion is of the order $\Delta t^3$ and can be neglected for small enough chosen time-steps. 
Usually the initial wave function $\psi(\mbfr, t)$ is given in position space and consequently the potential operator $\exp(-\rmi V \Delta t/2\hbar)$ is diagonal in this representation.
In order to apply the kinetic part $\exp(-\rmi T \Delta t/\hbar)$ a Fourier transform  $\cal F$ is used to express the wave function in momentum space. Thus the total time evolution of the wave packet is given by
\begin{eqnarray}
\psi(\mbfr,t+N\Delta t)
&\approx&\rme^{-\rmi V \Delta t/2\hbar}
{\left[{\cal F}^{-1}\rme^{-\rmi T \Delta t/\hbar}{\cal F}\rme^{-\rmi V \Delta t/\hbar}\right]}^N
\rme^{\rmi V \Delta t/2\hbar}\psi(\mbfr,t).
\end{eqnarray}
This scheme is easily adapted to various systems and gives converged results provided that the time-step $\Delta t$ is small enough \cite{Kramer2008a}.
Another possibility, especially suited for long time scales, is the direct expansion of the time-evolution operator in Chebyshev polynomials $C_n$ \cite{Kosloff1994a}
\begin{equation}
\psi (\mbfr,t+N\Delta t)  =  \sum_{n=0}^M (2-\delta_{n,0}) J_n(\Delta E N\Delta t/(2 \hbar))
\underbrace{C_n (- \rmi \hat{H}_\mathrm{norm}) \psi (\mbfr,t)  }_{P_n},
\end{equation}
where $J_n$ denotes the Bessel function of order $n$.
The order of the required polynomials $M$ is increased until $J_M(\Delta E N\Delta t/(2 \hbar))$ drops below the desired precision.
Since the Chebyshev polynomials form a perfect expansion of the exponential function on the interval $[-1,1]$, the propagation converges in the energy range $\Delta E$ if the Hamiltonian  is rescaled as $\hat H_\mathrm{norm}=2\hat\mathcal H/\Delta E$.
The states $P_n$ are obtained by the recursion relations
$P_0 =  \psi (\mbfr,t)$,
$P_1 =  \hat{H}_\mathrm{norm} \psi (\mbfr,t)$, and
$P_n =  - 2 \rmi \hat{H}_\mathrm{norm}  P_{n-1} +  P_{n-2}$.
The commutator error is absent in the polynomial expansion and long propagation times can be achieved.

In an open system the scattering region is coupled to asymptotic channels. One way to take into account the semi-infinite extent of the channels is to introduce a complex absorbing potential (CAP) $\rmi U(\mbfr)$ which is added to the potential $V(\mbfr)$. CAPs are widely used in quantum chemistry \cite{Jaeckle1996a}.
The complex-valued potential $U(\mbfr)$ is chosen to be zero in the scattering region and gradually grows in the channels towards the border of the simulation region (see fig.~\ref{fig:hallcross}(c)) in order to avoid spurious reflections. As a result the norm of the wave function decreases during the propagation.
The complex term $\rmi U(\mbfr)$ violates the self-adjointness of the Hamiltonian and leads to complex eigenvalues. Since the high-order Chebyshev expansion does not converge away from the real axis Faber or Newton polynomials have to be used for long-term propagations in open systems \cite{Kosloff1994a}.

\subsection*{Tracking the probability flux}

The transmission probabilities through the mesoscopic device is extracted from the time evolution of wave packets by
recording the probability flux in the asymptotic regions.
There are several possible ways to obtain the flux \cite{Tannor2007a}, (i) cross-correlation functions, (ii) fluxlines, or (iii) using complex absorber. The flux is recorded either time-resolved or energy resolved. Each approach has specific (dis)adventages and all of them can be used in conjunction. 
In the cross correlation approach we define a sender wave packet describing the initial situation. It is located in channel~$i$ populating
transversal mode $n_i$ and is composed of purely incoming momenta directed towards the scattering region. 
The destination, channel~$j$ with transversal mode $n_j$, is represented by an outgoing receiver state.
We record the overlap between forward propagated sender wave packet $\psi_{\rm send}(\mathbf{r},t)$
and stationary receiver wave packet $\psi_{\rm rec}(\mathbf{r})$ 
\begin{equation}
 C(t)=\langle \psi_{\mbox{\scriptsize rec}}| \psi_{\mbox{\scriptsize send}}(t) \rangle.
\end{equation}
The transmission amplitude is then given by the energy representation of the cross correlation function $C(t)$
\begin{equation}\label{eq:tcrosscorr}
t_{in_i\,jn_j}(E)=\frac{(2\pi\hbar)^{-1}}{\mu_{\mbox{\scriptsize rec}}^\ast(E)\,\eta_{\mbox{\scriptsize 
          send}}(E)}\int_{-\infty}^{\infty}C(t)\,\mbox{e}^{\mbox{\scriptsize{i}}Et/\hbar}\,\mbox{d}t.
\end{equation}
The factors $\eta_{\mbox{\scriptsize send}}$ and $\mu_{\mbox{\scriptsize rec}}$ 
are correction terms which account for the longitudinal shape of the wave packets and ensure that $t_{in_i\,jn_j}(E)$
does not depend on wave packet parameters. 
A second way consists in calculating the probability flux through a line in the asymptotic channels. The flux operator is given by
\begin{equation}
 \hat{F}=\frac{1}{2\,m}(\hat{p}\delta(\hat{x}-x_0)+\delta(\hat{x}-x_0)\hat{p}),
\end{equation}
where $x$ denotes the longitudinal direction of the channel and $x_0$ is the postion of the fluxline. 
The expectation value 
$\langle\psi|\hat{F}|\psi\rangle=\int \mathbf{j}(x_0,y) \rmd y$
gives the integrated flux along the fluxline.
The transmission propability is obtained by
\begin{equation}
 |t_{in_i\,jn_j}(E)|^2=\langle\psi_{j,n_j}^+(E)|\hat{P}_{i,n_i} \hat{F}\hat{P}_{i,n_i}|\psi_{j,n_j}^+(E)\rangle,
\end{equation}
where
\begin{equation}
 \hat{P}_{i,n_i}=\int\,|\psi_{i,n_i}^-(E')\rangle\langle\psi_{i,n_i}^-(E')|\,\rmd E'
\end{equation}
projects onto outgoing states in channel $i$ with transversal mode $n_i$. Since the fluxline is located in the asymptotic channel we can 
relate the scattering eigenstate $|\psi_{i,n_i}^-(E)\rangle$ to channel eigenstates $\psi_{\alpha,n,\pm}=\chi_{\alpha,n,\pm k_x}\,e^{\pm \mbox{\scriptsize i}\, k_x x_\alpha}$.
The scattering eigenstate
\begin{equation}
 |\psi_{j,n_j}^+(E)\rangle =\frac{(2\pi\hbar)^{-1}}{\eta_{\mbox{\scriptsize send}}(E)}\int_{-\infty}^\infty\ \psi_{\mbox{\scriptsize send}}(t)
 e^{\mbox{\scriptsize{i}} Et/\hbar}\,\mbox{d}t
\end{equation}
is extracted from the propagation history of the sender wave packet.
The third method works by calculating the reduction of the norm due to presence of the CAP within each channel and to obtain the flux from the change of the norm between successively applied absorption events in each channel.

\section{Application example: Hall cross}
\begin{figure}[t]
\includegraphics[width=0.65\textwidth]{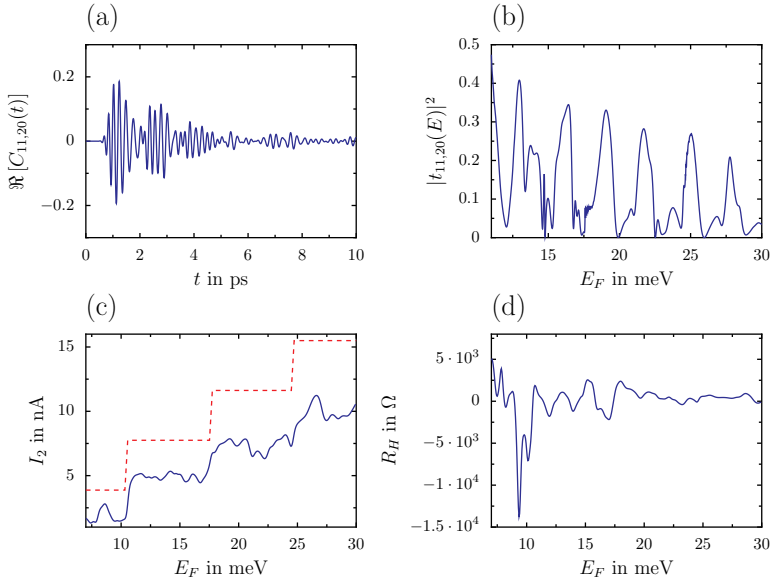}
\caption{Transform of the time-dependent cross-correlation to the energy-resolved transmission for the Hall cross in fig.~\ref{fig:hallcross} at a magnetic field of ${\cal B}=0.4$~T. (a) Time-dependent cross-correlation function for the transmission of lead 2, mode 0 into arm 1, mode 1. (b) Transmission amplitude, obtained by the Fourier transform of (a). (c) Current $I$ through the system as function of Fermi energy for a voltage of $0.1$~mV and temperature 1~K. The dashed line corresponds to the step-wise changes of the current in the harmonic waveguide, eq.~(\ref{eq:jnet}). (d) Resistivity as function of the Fermi energy. \label{fig:transmission}}
\end{figure}
As an example of the formalism, we discuss the transmission through the device shown in fig.~\ref{fig:hallcross}, which is a four terminal device and can be used to probe the Hall effect. For a numerical evaluation, several practical constraints have to be fulfilled: the wave packet must fit on the numerical grid in configuration and in momentum space. In order to use the fast Fourier transform (FFT) it is desirable to use the same number of grid points in both representations \cite{Feit1982a}. 
The discretization of the position and momentum space gives a lower limit for the spatial extension of the used wave packets, since one has to ensure that each packet is sampled sufficiently.
And last, numerically we cannot propagate the wave packets infinitely long so that we have to stop the time evolution after a certain time. The propagation time is chosen long enough that most parts of the wave packet already traveled through the device and got absorbed in the CAPs at the end of the channels. In fig.~\ref{fig:transmission}(a) we illustrate the detected overlap between receiver and sender wave packets. 
The sender was chosen to represent the transversal ground state in lead~2 and the receiver populates 
the first exited mode in arm~1. During time evolution the sender packet propagates through the Hall cross, scatters in different channels, and finally gets absorbed. Hence the cross correlation function decays with time.
After 120~ps the absolute value of the cross correlation remains below $5\times10^{-5}$ and we stop the propagation. From eq.~(\ref{eq:tcrosscorr}), the transmission propability is obtained by performing the Fourier transform of the cross correlation function which is done efficiently with a FFT. We get the transmissions for discrete energy values where the spacing is determined by
the total propagation time. For the present simulation we got 116 energy points per meV. Fig.~\ref{fig:transmission}(b) shows the
propability for inter mode scattering from lead~2 with mode~0 to arm~1 in mode~1.\\
In contrast to quantum chemistry where the wave packet initially represents an eigenstate of a certain 
potential energy surface, wave packets are not associated with a direct physical meaning in their application for mesoscopic transport. Rather they are used as tools to compute transmission probabilities for a scattering potential and we can match the longitudinal shape to the needs at hand. It is convenient to use Gaussian wave packets whose parameters are adjusted to represent a certain energy range. A single wave packet run gives then the transmission amplitudes for an energy range which is sufficiently represented by the used wave packet. To increase the numerical efficiency we set the Gaussian width as small as possible to get an extended energy representation.  A further enhancement of efficiency is achieved by using several receiver states within a single propagation run. Here, we put five receiver states in each lead populating the transversal modes up to $n=4$. Thus we obtain about 40,000 transmission amplitudes corresponding to a typical computation time of 0.2~s per amplitude on a standard CPU. Current and voltages in the four terminal Hall cross are evaluated by inserting the computed transmission amplitudes in eq.~(\ref{eq:current}). We apply a bias voltage of 0.1~mV which drops symmetrically between contact~2 and contact~3 around the Fermi energy $E_F$. The temperature is set to $1$~K corresponding to $4k_BT=0.35$~meV. Contacts 1 and 4 act as perfect voltage probes forcing the currents $I_1$ and $I_4$ to vanish. This gives a nonlinear system of equations, whose solution determines the chemical potentials $\mu_1$ and $\mu_4$ and hence the voltage drop $V_H=(\mu_1-\mu_4)/e$. In fig.~\ref{fig:transmission}(c) we show the current between lead~2 and lead~3 in dependency on the Fermi energy. The conductance quantization of the harmonically confined channel gives rise to a step like behavior of the current. Note that the current remains below the conductance quantization of a harmonic waveguide (dashed line) since electron transmission in the side arms as well as reflections diminish the conductance through the Hall cross. Furthermore the non-zero temperature smears out structures on the energy scale of $4k_BT$ and especially any abrupt jumps at mode openings get broadened.
Fig.~\ref{fig:transmission}(d) shows the Hall resistance $R_H=V_H/I_2$ which goes down with increasing number of open modes. Since single mode effects superpose each other, $R_H=V_H/I_2$ reflects an averaged behavior and gets more regular with increasing Fermi energies.

\section{Open questions in mesoscopic physics}

The time-dependent approach to mesoscopic physics can also be generalized to materials with more complicated band-structures. In general one has to propagate wave packets with additional spin or pseudo-spin degrees of freedom, like the two-dimensional time-dependent Dirac equation for graphene \cite{Krueckl2009a}.

The previous calculation assumed an effective non-interacting description of the electrons in the nanodevice. The inclusion of interactions beyond the mean-field level is a complicated task, but is required in order to explain many observed phenomena. As an example we mention the quantum Hall effect, where the existence of the Hall potential is a direct consequence of interaction effects \cite{Kramer2009c,Kramer2009b}. In principle the wave packet formalism can be extended to incorporate interactions between electrons, but this generalization requires to propagate a properly (anti-)symmetrized product of single-particle orbitals. A more serious conceptional problem arises due to the injection of electrons into the device, which requires to change the device many-body electronic wave function. Most current approaches switch off the interactions away from the scattering region. However, this procedure may not be compatible with boundary conditions at the contacts. Self-consistent schemes are under development but require further work in order to explain the experimental observations \cite{Albareda2009a,Kramer2009c,Kramer2009b}.

\ack

T.K. would like to thank V.~Dodonov for the kind invitation to present this work at the conference Quantum Non-stationary Systems in Brasilia, Brazil, and R.~E.~Parrott, E.~J.~Heller, and U.~Kunze for helpful discussions. We appreciate funding by the Emmy-Noether program of the DFG, grant KR 2889/2.

\section*{References}

\providecommand{\url}[1]{#1}

\end{document}